\documentstyle[psfig,floats,aps,twocolumn]{revtex}

\begin{document}
\draft

\twocolumn[\hsize\textwidth\columnwidth\hsize\csname 
@twocolumnfalse\endcsname

\title{Phase-field modeling of microstructural
pattern formation during directional 
solidification of peritectic alloys without morphological instability}

\author{Tak Shing Lo$^{1,2}$, Alain Karma$^1$, and Mathis Plapp$^{1,3}$}

\address{$^1$Physics Department and Center for Interdisciplinary
                Research on Complex Systems \\
                 Northeastern University, Boston MA 02115 \\
         $^2$Courant Institute of Mathematical Sciences \\
             New York University, New York NY 10012 \\
         $^3$Laboratoire de Physique de la Mati{\`e}re Condens{\'e}e\\
             CNRS/Ecole Polytechnique, 91128 Palaiseau, France
}

\date{April 15, 20001}

\maketitle

\begin{abstract}

During the directional solidification 
of peritectic alloys, two stable solid phases (parent and peritectic)
grow competitively into a metastable liquid phase of larger
impurity content than either solid phase. When the parent or
both solid phases are morphologically unstable,
i.e., for a small temperature gradient/growth rate
ratio ($G/v_p$), one solid phase usually outgrows and covers the
other phase, leading to a cellular-dendritic array structure 
closely analogous to the one formed during monophase solidification of a
dilute binary alloy. In contrast, 
when $G/v_p$ is large enough for both phases to be 
morphologically stable, the formation of the microstructure 
becomes controlled by a subtle interplay between the
nucleation and growth of the two solid phases.  
The structures that have been observed in this regime 
(in small samples where convection effects
are suppressed) include alternate 
layers (bands) of the parent and peritectic phases perpendicular 
to the growth direction, which are formed by alternate 
nucleation and lateral spreading of one phase
onto the other as proposed in a recent model {\rm [}R. Trivedi, 
Metall. Mater. Trans. A {\bf 26}, 1 (1995){\rm ]},
as well as partially filled bands (islands), where
the peritectic phase does not fully cover the
parent phase which grows continuously. We develop a  
phase-field model of peritectic solidification that
incorporates nucleation processes in order to explore the formation of
these structures. Simulations of this model shed light on the
morphology transition from islands to bands, the dynamics of spreading
of the peritectic phase on the parent phase following nucleation,
which turns out to be characterized by a remarkably constant
acceleration, and the types of growth morphology that one might expect
to observe in large samples under purely diffusive growth conditions.
\end{abstract}
]

\section{Introduction}

The spontaneous emergence of complex microstructural patterns 
during the solidification of alloys is a subject of both 
fundamental and applied interest \cite{review}. 
During directional solidification, a sample is pulled
in an externally imposed temperature gradient $G$ with a
fixed pulling speed $v_p$. This setup has been used extensively
in fundamental studies of solidification patterns
because it allows one to study their
formation under well-controlled growth conditions. Depending
on the type of alloy and the ratio $G/v_p$, various 
patterns are possible. During monophase solidification of
a dilute binary alloy, solute redistribution
leads to a well-known morphological instability 
(Mullins-Sekerka instability \cite{MulSek}) below a critical ratio 
$G/v_p$, and cellular or dendritic patterns are typically formed.
For nondilute alloy concentrations close to a
eutectic point, two stable solid phases of different compositions
can grow from a metastable liquid. In this case, the two phases
cooperate and form lamellae or rods parallel to the growth
direction (coupled growth). For off-eutectic compositions, 
coexistence between dendrites and coupled growth structures 
is also observed.

\begin{figure}
\centerline{
  \psfig{file=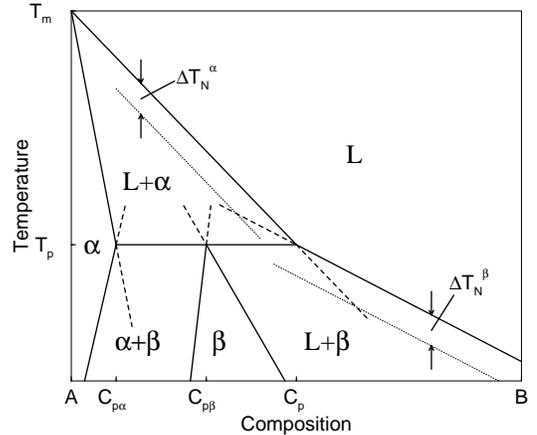,angle=-90,width=.45\textwidth}}
\caption{Schematic phase diagram of a peritectic alloy. 
$C$, concentration of impurity B; $T_m$, melting point of pure A;
$T_p$, peritectic temperature. $C_p$, $C_{p\beta}$, and $C_{p\alpha}$
are the compositions of the liquid, $\beta$ solid, and $\alpha$ solid
that are in equilibrium at $T_p$.
$\Delta T_N^{\alpha}$ and $\Delta T_N^{\beta}$ are the nucleation 
undercoolings for $\alpha$ and $\beta$ phases, respectively. Dashed lines
are metastable extensions of the liquidus and solidus lines.}
\label{schematic}
\end{figure}

Much less is known about microstructural
pattern formation in peritectic growth \cite{review}, despite the
fact that many industrially important metallic alloy systems 
as well as ceramics such as the high-$T_c$ superconductor 
YBCO are peritectics. A schematic phase diagram of a peritectic  
AB-alloy (where B will be called the impurity
for convenience) is shown in Fig.~\ref{schematic}.
It contains a peritectic point, analogous
to the eutectic point, at which two different solid phases, the 
parent (primary) and peritectic (secondary) phases, coexist
with a liquid of higher composition than either solid phase.
Above the peritectic temperature $T_p$, the parent phase 
is stable and the peritectic phase is metastable, whereas 
below $T_p$, the opposite is true. For comparison, in a eutectic,
both solid phases are stable below the eutectic temperature, 
and metastable above, and the impurity concentration in the
liquid falls in between the concentrations of the two solid phases.
For a sufficiently low $G/v_p$ ratio, a dendritic array
structure of the parent {\it or} the peritectic phase is typically observed, and
which of these two phases is selected depends on the alloy composition and $G/v_p$ \cite{Umeetal}.  
In contrast, for a high $G/v_p$ ratio morphological
instability is suppressed. In this case, banded structures made up of 
alternating layers of primary and peritectic phases 
perpendicular to the growth direction are formed. These
structures have by now been observed in 
various peritectic systems, including Sn-Cd \cite{Boettinger74,Brody77}, 
Sn-Sb \cite{Titchener75}, Zn-Cu \cite{Titchener75}, 
Ag-Zn \cite{Ostrowski77}, and Pb-Bi \cite{Barker74,Fuh84}.
It is worth noting that eutecticlike coupled 
growth structures, which are quite distinct from banded
structures, have recently been observed in the Fe-Ni system \cite{VandPhD}.
Whether stable coupled growth is theoretically possible during
peritectic growth has remained an open question for quite some time \cite{Boettinger74},
and we will address this issue elsewhere. Here, we focus primarily on
banded structure formation and phenomena associated with
the dynamical spreading of one solid phase onto the other.

Recently, Trivedi has introduced a one-dimensional (1D) model  
\cite{Trivedi95} to explain the formation of peritectic banded
structures for purely diffusion-controlled growth 
The conceptual banding cycle assumed in this model is as follows. 
Consider a melt with homogeneous composition $C_{\infty}<C_p$ being solidified 
starting from a flat $\alpha$-liquid interface in equilibrium.
The rejection of impurities B into the liquid during solidification
leads to the buildup of a solutal boundary layer. As a result,
the interface temperature decreases, following the liquidus curve
in the phase diagram. If $C_{\infty}$ is large enough, 
the interface temperature eventually falls sufficiently below $T_p$ for
the peritectic phase to nucleate heterogeneously at the solid-liquid interface
before the growth of the $\alpha$ phase has reached its steady state. The
newly nucleated $\beta$ phase rejects fewer impurities than the
$\alpha$ phase. Consequently, the magnitude of the solutal boundary layer
decreases and the interface temperature increases, following now
the $\beta$-liquid coexistence line in the phase diagram.
If $C_{\infty}$ is low enough, such that the corresponding
interfacial temperature is sufficiently higher than $T_p$, the $\alpha$ solid may renucleate
again before the steady state is reached, and the cycle repeats.  
Therefore, this model predicts that bands can form only
when the composition falls inside a narrow window in the hypoperitectic 
region ($C_{p\alpha}<C_{\infty}<C_{p\beta}$) whose width depends on
the nucleation undercoolings  $\Delta T_N^{\alpha}$ 
and $\Delta T_N^{\beta}$.

The first attempts to validate 
this prediction experimentally yielded contradictory results.  
Directional solidification experiments 
with Pb-Bi and Sn-Cd alloys seemed to show that
bands also form in the hyperperitectic region
($C_{p\beta}<C_{\infty}<C_p$), in
apparent contradiction with this prediction.
An attempt was made to resolve this ``composition 
range paradox'' by incorporating convection effects \cite{Karma98a}, 
assuming the existence of a fully mixed liquid of uniform composition 
outside a purely diffusive 1D boundary layer of finite thickness.
This model, however, yielded a banding cycle and band spacings  
that are inconsistent with experimental
results, hinting that this boundary-layer approximation (typically valid
for strong convection) is inadequate
to describe these experiments. Around the same time, careful serial sectioning
of solidified Pb-Bi and Sn-Cd alloys revealed that the
seemingly banded structures are actually oscillatory treelike
structures connected in three dimensions 
\cite{Park98}, and not discrete bands, thereby resolving experimentally
this composition range paradox. Following this finding, 
a more accurate model was developed that assumes a planar solidification front,
but incorporates a fully two-dimensional convection flow field 
\cite{Mazumder98}. This model successfully reproduced the
observed oscillatory structures.  

Following these studies, experiments were
conducted in thin tubes to reduce convection \cite{Park98}. For
tube diameters smaller than 1 mm, truly discrete bands 
indeed became observable inside a narrow composition
range predicted by the 1D diffusive growth model.
Surprisingly, however, it was also
observed that when the tube diameter was further reduced, 
``islands'' of the $\beta$ phase formed inside the matrix of 
the $\alpha$ phase, instead of discrete bands. 
This observation suggests that there is a microstructural transition 
from bands to islands if the system size is reduced. It was also observed 
that islands tend to form more easily for initial compositions closer to
$C_{p\alpha}$. In addition, some spatially chaotic patterns were 
observed in some experiments. The formation of these structures is controlled by a
subtle interplay between the nucleation process and the competition between the growth of
the nuclei and the preexisting phase. In this respect, the one-dimensional
model may not always be adequate to describe
this competition because it assumes an infinite spreading speed for the
newly nucleated phase. Moreover, the 2D convection model
assumes a flat interface and is hence not well suited to simulate heterogeneous
nucleation and spreading. In order to model accurately the formation of these
different structures, a truly 2D model of interface evolution is necessary. 
The particular difficulty of this problem is that the microstructure
formation is controlled by an interplay between nucleation and growth
of the different phases. No steady-state growth mode exists, 
which makes the whole problem explicitly time-dependent.

In this paper, we use a phase-field 
approach \cite{Langer86,Wheeler92,Warren95,Karma98b,Kop98,Karma94,Elder94,Wheeler96}
to investigate the formation of this class of banded microstructures in a
purely diffusive regime and a 2D geometry.
The phase-field method eliminates the need of explicit front
tracking and thus greatly simplifies the task of numerically solving the
equations of peritectic solidification that involve three-phase junctions. 
A phase-field model for peritectic
growth has recently been proposed \cite{Steinbach98}. Here, we use an
alternative model that is closer to the eutectic model
of Wheeler {\em et al.} \cite{Wheeler96}.  

We first investigate the spreading of the peritectic
phase on the primary phase after a single nucleation event. 
We characterize in detail the dynamics of the three-phase junction
during spreading and find a morphological transition from discrete bands of $\alpha$ and $\beta$
phases to isolated islands of $\beta$ phase when the system size is
decreased, in qualitative agreement with experiments. Moreover,
our simulations enable us to understand physically the basic mechanism
that underlies this transition. We then investigate the
effect of multiple nucleations on microstructure formation
in large systems by supplementing the phase-field equations 
with a phenomenological stochastic nucleation law.  

The remainder of this article is organized as follows. In Sec. II, we
write down the sharp interface and phase-field models. Section III is 
devoted to the study of the equilibrium properties of the phase-field model and Sec. IV
describes the simulation method. Results are presented in 
Sec. V, followed by a summary and conclusions in Sec. VI. 

\section{Model}
\subsection{Sharp-interface model}

The sharp-interface equations are given by
\begin{equation}
\partial_t C=D_L\nabla^2 C, \label{diffeqt}
\end{equation}
\begin{equation}
v_n(C_L-C_{\nu})=-D_L \partial_n C_L, \label{masscon}
\end{equation}
\begin{equation}
T=T_p+m_{\nu}(C_L-C_p)-\Gamma_{\nu}K-\frac{1}{\mu_{\nu}}v_n,  \label{GTR}
\end{equation}
where $C$ denotes the concentration of impurity B, and the subscript $\nu$
labels the solid $\alpha$ and $\beta$ phases. Equation~(\ref{diffeqt}) is the
diffusion equation for the solute in the liquid with the solute
diffusivity $D_L$. We have assumed that diffusion in the solid is
negligible (one-sided model). Equation~(\ref{masscon}) expresses the mass
conservation at the moving interface, with $v_n$ and $\partial_n$ denoting
the normal velocity of the interface and the derivative normal to the
interface, respectively. Finally, Eq.~(\ref{GTR}) is the Gibbs-Thomson
condition at the solid-liquid interface, with $K$, $m_{\nu}$, $\mu_{\nu}$,
and $\Gamma_{\nu}$ being the interface curvature, liquidus slope, kinetic
coefficient, and Gibbs-Thomson constant of phase $\nu$, respectively.  The
Gibbs-Thomson constants $\Gamma_{\nu}$ are defined by
\begin{equation}
\Gamma_{\nu}=\frac{\gamma_{\nu L}T_p}{L_{\nu}},
\end{equation}
where $\gamma_{\nu L}$ is the surface energy of the $\nu$-liquid interface 
and $L_{\nu}$ is the latent heat of fusion for phase 
$\nu$, both taken at the peritectic temperature. 
Young's condition 
\begin{equation}
\gamma_{\alpha L}{\bf t}_{\alpha L}+\gamma_{\beta L}{\bf t}_{\beta L}+
\gamma_{\alpha\beta}{\bf t}_{\alpha\beta}=0,        \label{Young}
\end{equation}
must be satisfied at the 
trijunction points where three phases meet,
where ${\bf t}_{\mu\nu}$ is the unit vector
parallel to the $\mu$-$\nu$ interface and pointing away from the
trijunction.  

\subsection{Phase-field model}

To distinguish between the three
possible phases (liquid, $\alpha$ solid, and $\beta$ solid), we follow a
similar approach to that of Wheeler {\it et al.} \cite{Wheeler96} for eutectic
solidification by introducing two nonconserved order parameters 
(phase fields) $\phi$ and $\psi$. The first distinguishes between solid
($\phi=1$) and liquid ($\phi=-1$), the second between the
$\alpha$ solid ($\psi=1$) and the $\beta$ solid ($\psi=-1$). 
The solid-liquid interface is defined by the level curve $\phi=0$, 
and the interface between the solid $\alpha$ and $\beta$ 
phases is defined by the level curve $\psi=0$ when $\phi$ is positive. 
One important difference
from Ref. \cite{Wheeler96} is that in our model $\psi$ takes the
well-defined value $\psi=0$ in the liquid. This modification is necessary
because, in the model of Wheeler {\it et al.}, the equation of motion for
$\psi$ becomes a simple diffusion equation in the liquid. This introduces
an undesirable new time scale in dynamical simulations that is removed
in the present approach. 

As a third dynamical variable we need the composition $C$, 
which is a conserved field. We define the scaled composition
\begin{equation}
c({\bf r},t)=[C({\bf r},t)-C_{p\beta}]/\Delta C_{\alpha},
\label{cscale}
\end{equation}
where $\Delta C_{\nu}=(C_p- C_{p\nu})$, $\nu=\alpha,\beta$, is
the concentration jump at the $\nu$-liquid interface at $T_p$.

In terms of these quantities, the equations of motion that govern the
dynamics of the system are given by
\begin{eqnarray}
\tau_\phi\frac{\partial \phi}{\partial t}&=&-\frac{\delta F}{\delta\phi},
                                                    \label{EM1}       \\
\tau_\psi\frac{\partial \psi}{\partial t}&=&-\frac{\delta F}{\delta\psi},
                                                    \label{EM2}       \\
\frac{\partial c}{\partial t}&=&\nabla\cdot\left[M(\phi)
    \nabla{\delta F\over \delta c}\right],
                                                    \label{EM3}
\end{eqnarray}
where $F$ is the dimensionless free energy of the system
(i.e., the Helmholtz free energy, divided by the product
of the system size and a typical value of the free energy
density that sets the physical energy scale),
$M(\phi)$ is the mobility of the impurities, and
$\tau_\phi$ and $\tau_\psi$ are (fast) relaxation times
for the phase fields. These equations are of the 
standard variational form known from out-of-equilibrium 
thermodynamics. Note that, since $\delta F/\delta c$ is the
local chemical potential $\mu$, Eq.~(\ref{EM3}) is simply 
the continuity equation for the impurity concentration
with the mass current ${\bf J}$ given by
\begin{equation}
{\bf J}=-M(\phi)\nabla\mu.
\end{equation}
If there are no fluxes across the boundary of the volume 
where $F$ is defined,  $dF/dt\le 0$ and Eqs.~(\ref{EM1})-(\ref{EM3}) imply that
the dynamics drives the system toward 
a minimum of free energy.  

The free energy functional of the system is assumed to 
be of the form
\begin{equation}
F=\int \{\frac{1}{2}W^2_\phi|\nabla\phi|^2
   +\frac{1}{2}W^2_\psi|\nabla\psi|^2+f(\phi,\psi,c)\}d^3{\bf r}.
\end{equation}
Since $F$, $\phi$, and $\psi$ are dimensionless, the coefficients 
$W_\phi$ and $W_\psi$ have the dimension of length: they determine 
the width of the diffuse interfaces.
The form of the free energy density is chosen such that 
there are two minima at $\psi=\pm 1$ corresponding to the   
$\alpha$ ($+$) and $\beta$ ($-$) phases
for $\phi=+1$. There is a single minimum in
the liquid corresponding to $\phi=-1$ 
and $\psi=0$, and $f(\phi,\psi,c)$ 
has a single minimum as a function of $c$
for fixed values of $\phi$ and $\psi$ corresponding 
to the three equilibrium phases.
A convenient way to match these requirements is to construct a
free energy density of the form
\begin{eqnarray}
f(\phi,\psi,c)&=&\frac{\lambda}{2}\bigl\{c+A_1h(\phi)+\frac{1}{2}A_2[1+h(\phi)]
                      h(\psi)\bigr\}^2 \nonumber \\
              & & -\lambda\bigl\{B_1h(\phi)+\frac{1}{2}B_2
                   [1+h(\phi)]h(\psi)\bigr\}+g(\phi) \nonumber \\
              & & +\frac{1}{2}\bigl[1+h(\phi)\bigr]g(\psi)
               +\frac{1}{2}\bigl[1-h(\phi)\bigr]\psi^2. \label{freeenden}
\end{eqnarray}
Here, $\lambda$ is a positive constant, and $A_1$, $A_2$, $B_1$, and
$B_2$ are functions of temperature. The function $g$ is a
double-well potential with minima at $\pm 1$, and the function $h$
must satisfy $h(\pm1)=\pm1$ and $h'(\pm 1)=0$ in order to keep the
minima of $f$ at constant values of $\phi$ and $\psi$,
independent of the value of $c$. We take
\begin{eqnarray}
g(\phi)&=&1/4-\phi^2/2+\phi^4/4, \label{gphi} \\
h(\phi)&=&3(\phi-\phi^3/3)/2. \label{hphi}
\end{eqnarray}
The functions $g(\psi)$ and $h(\psi)$ are similarly defined. It follows trivially from
Eq.~(\ref{freeenden}) that the bulk phase free energy densities are given by
\begin{eqnarray}
       f_L&\equiv& f(-1,0,c)= \frac{\lambda}{2}(c-A_1)^2+\lambda B_1,
                                                \label{fL}    \\
f_{\alpha}&\equiv& f(1,1,c)=\frac{\lambda}{2}(c+A_1+A_2)^2-\lambda(B_1+B_2), 
                                                \label{fsalp} \\
 f_{\beta}&\equiv& f(1,-1,c)=\frac{\lambda}{2}(c+A_1-A_2)^2-\lambda(B_1-B_2).
                                                \label{fsbet}
\end{eqnarray}
For the mobility function $M(\phi)$, we take
\begin{equation}
M(\phi)=\frac{D_L}{2 \lambda}(1-\phi).
\end{equation}
With this choice, the diffusion coefficient of the impurity is a
constant equal to $D_L$ in the liquid and zero in both solids, which
corresponds to the so-called one-sided model.  
A standard asymptotic analysis of the sharp-interface limit of the
present phase-field model \cite{lothesis} 
shows that Eqs.~(\ref{EM1})--(\ref{EM3}) reduce as expected to
Eqs.~(\ref{diffeqt})--(\ref{GTR}). The relation between
the parameters in the two sets of equations is given in 
the next section.

\section{Phase diagram and Equilibrium properties}
 
By applying the well-known common tangent construction to the 
bulk free energy densities given by Eqs. (\ref{fL})--(\ref{fsbet}), we
can construct the equilibrium phase diagram of the phase-field model.
The equilibrium compositions can be expressed in terms of
the temperature-dependent 
functions $A_1$, $A_2$, $B_1$ and $B_2$ (see the appendix). However, since there are
only four functions, we can at most fit four lines out of six in the phase
diagram (i.e., three pairs corresponding to $\alpha$-liquid, $\beta$-liquid,
and $\alpha$-$\beta$ coexistence). We may choose to construct, say, the two liquidus lines and the two
solidus lines and leave the two solid-solid coexistence curves determined by
Eq.~(\ref{SS1}) and Eq.~(\ref{SS2}). Since we are interested only in the 
behavior of the system at temperatures close to $T_p$, we assume for 
simplicity that the liquidus and solidus lines are straight, and that the 
concentration jumps at the solid-liquid interface are constant (liquidus
and solidus are parallel). We can then choose $A_1$ and $A_2$ as constants 
and $B_1$ and $B_2$ as linear functions of the temperature. 
The corresponding expressions for the functions $A_1$, $A_2$, $B_1$, and $B_2$
are given in the appendix expressed in terms of the dimensionless temperature field
\begin{equation}
{\tilde T}=\frac{(T-T_p)}{|m_{\alpha}|\Delta C_{\alpha}}, \label{scaledtemp}
\end{equation}
which is a measure of the temperature relative to $T_p$ normalized by 
the freezing range of the $\alpha$ phase.

\begin{figure}
\centerline{
 \psfig{file=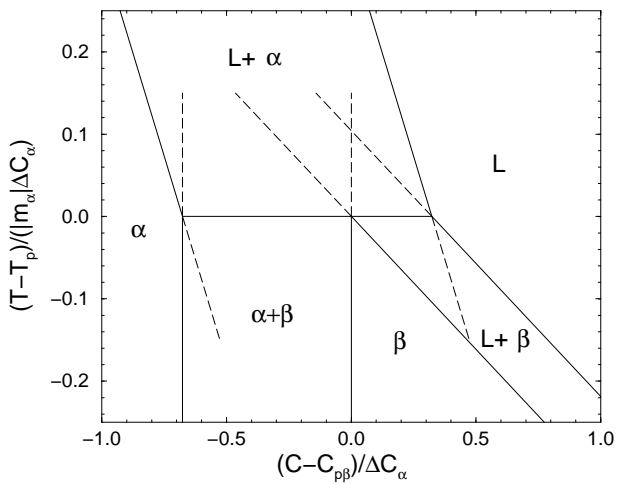,angle=-90,width=.45\textwidth}}
\smallskip
\caption{Phase diagram for our model system. Dashed lines
are metastable extensions of the liquidus and solidus lines.
}\label{phase_dia}
\end{figure}

In the present model, there exists a temperature-dependent
concentration $c_u$ such that, 
in the solid, the solid $\alpha$ ($\beta$) phase is thermodynamically
stable only if $c<c_u$ ($c>c_u$). By comparing Eq.~(\ref{fsalp}) and
Eq.~(\ref{fsbet}), it is easy to show that $c_u$ is exactly midway
between the two solid-solid coexistence lines. In order to avoid a
phase transformation in the solid far behind the solid-liquid interface,
we require $c_u$ to be independent of temperature. One
way to achieve this is to make the solid-solid coexistence lines vertical 
by choosing suitable parameters. This difference from a real peritectic phase
diagram is unlikely to change the qualitative behavior of the system. A
phase diagram for the model system used in our simulations is shown in 
Fig.~\ref{phase_dia}. 

\begin{figure}
\centerline{
 \psfig{file=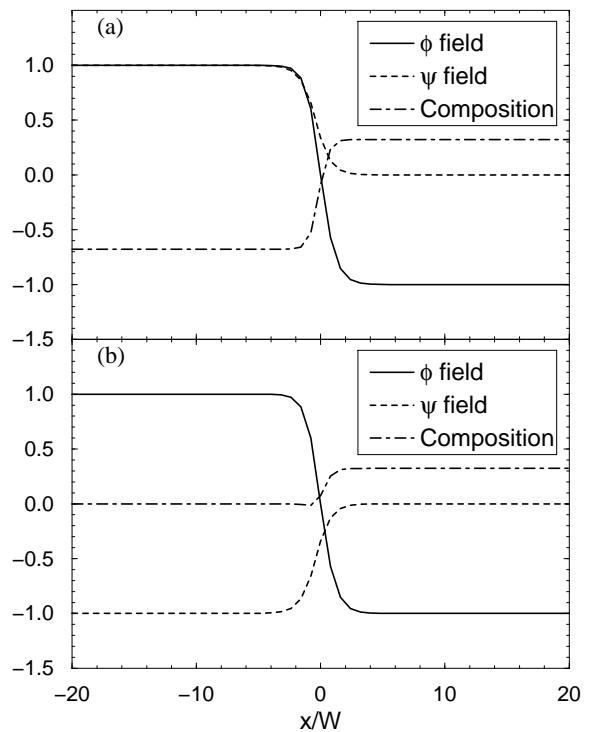,angle=180,width=.45\textwidth}}
\caption{Equilibrium profiles for (a) $\alpha$-L equilibrium
and (b) $\beta$-L equilibrium at $T_p$ ($\lambda=2.5$). The 
composition is scaled according to Eq. (\ref{cscale}).
}\label{eqmprofile}
\end{figure}

The equilibrium interface profiles connecting different phases can be
obtained by solving the time-independent one-dimensional version of the
equations of motion with suitable boundary conditions. Since the chemical
potential must be constant at equilibrium, the relation
\begin{equation}
\mu=\frac{\delta F}{\delta c}=\frac{\partial f}{\partial c} \label{chempo}
\end{equation}
can be used to eliminate the concentration field from Eqs.~(\ref{EM1}) and
(\ref{EM2}). The appropriate value of $\mu$ for a certain temperature
is obtained from the common tangent construction. The two resulting
coupled ordinary differential equations were solved numerically using a
Newton-Raphson method on a one-dimensional grid of spacing $\Delta x$. For
simplicity, we assumed $W_\phi=W_\psi=W$. Unless otherwise stated,
all the results below are obtained for $\Delta x/W=0.8$, which
provides a good compromise between computational efficiency and
accuracy. The resulting equilibrium
profiles, centered at the origin, for the phase fields and the
concentration for $\alpha$-L equilibrium and $\beta$-L equilibrium at
$T_p$ are shown in Fig.~\ref{eqmprofile}. For solid-solid equilibrium, the
interface profile of $\psi$ can be obtained analytically because $\phi=+1$ is a
constant:
\begin{equation}
\psi_0(x)=-\tanh(\frac{x}{\sqrt{2}W}).
\end{equation}
In all cases, the concentration profiles are given by substituting the
equilibrium profiles $\phi_0(x)$ and $\psi_0(x)$ obtained previously into
Eq.~(\ref{chempo}).
With the equilibrium interface profiles at hand, we can calculate the
surface energies $\gamma_{\alpha L}$, $\gamma_{\beta L}$, and
$\gamma_{\alpha\beta}$, defined as the excess Gibbs free energy per unit
surface area. They are given by the expressions
\begin{equation}
\gamma_{\mu\nu}=
\int_{-\infty}^\infty [ W_\phi^2 (\partial_x \phi_0)^2
     +W_\psi^2 (\partial_x \psi_0)^2] dx,           \label{SE1}
\end{equation}
where $\phi_0$ and $\psi_0$ are the equilibrium profiles of the
phase fields connecting phases $\mu$ and $\nu$. The same formula 
for the surface energies can also be obtained by a
matched asymptotic expansion \cite{lothesis}.
For the solid-liquid interfaces, the surface energies are obtained by
numerical integration. For this purpose, it is more convenient to
convert Eq.~(\ref{SE1}) to a form without the gradients of the fields.
Making use of the steady-state one-dimensional 
version of the equations of motion and the fact that $\mu$ is 
constant in equilibrium, we obtain after some algebra 
\begin{equation}
\frac{d}{dx}\left[\frac{1}{2}[W_\phi^2 (\partial_x\phi_0)^2+W_\psi^2
(\partial_x \psi_0)^2]\right]=\frac{d}{dx}(f-\mu c).  \label{SEinter}
\end{equation}
Now one can integrate Eq.~(\ref{SEinter}) from $-\infty$ to an
arbitrary $x$ and make use of the expression for the equilibrium
concentration profile and the bulk phase values to show that for the  
solid-liquid interfaces,
\begin{eqnarray}
 & &
\frac{1}{2}[W_\phi^2 (\partial_x\phi_0)^2+W_\psi^2 (\partial_x \psi_0)^2]
                                                \nonumber \\
 &=& 
g(\phi_0)+\frac{1}{2}[1+h(\phi_0)]g(\psi_0)+\frac{1}{2}[1-h(\phi_0)]\psi_0^2
                                                \nonumber \\
 & &
 +\frac{\lambda}{2}\left(\frac{\bar{B}}{\bar{A}}-B_2\right)[1+h(\phi_0)]
       [h(\psi_0)\mp1].                             \label{SE2}
\end{eqnarray}
Here the upper and lower signs are for $\alpha$-liquid and $\beta$-liquid
equilibrium, respectively, and $\bar{A}$ and $\bar{B}$ are defined in
the appendix. For the solid-solid interface, the surface energy
$\gamma_{\alpha\beta}$ can be calculated exactly and is equal to
$2\sqrt{2}W_\psi/3$. 

\begin{figure}
\centerline{
 \psfig{file=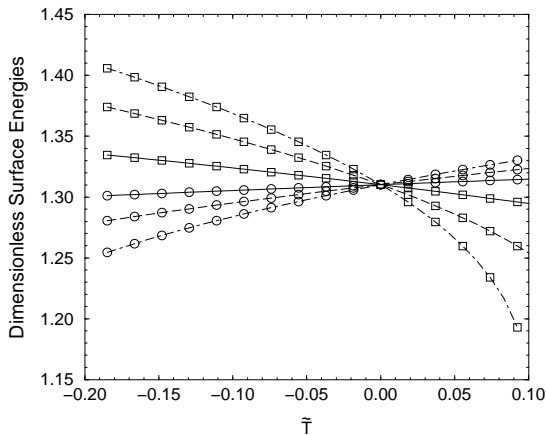,angle=-90,width=.4\textwidth}}
\caption{Dimensionless solid-liquid surface energies versus temperature
for different $\lambda$. Circles, $\gamma_{\alpha L}/W$; squares,
$\gamma_{\beta L}/W$. Solid lines, $\lambda=0.5$; dashed lines,
$\lambda=1.5$; and dash-dotted lines, $\lambda=2.5$.}    \label{surten}
\end{figure}

\begin{figure}
\centerline{
  \psfig{file=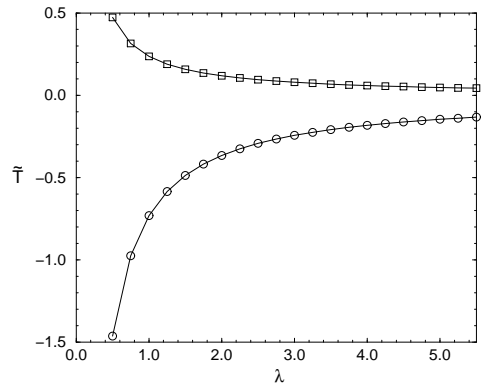,angle=-90,width=.4\textwidth}}
\caption{Temperature range within which both $\alpha$-liquid and
$\beta$-liquid equilibrium solutions exist, versus $\lambda$. 
(Circles: limit of existence for $\alpha$-liquid interface solution. 
Squares: limit of existence for $\beta$-liquid solution.)}                   \label{T_limit}
\end{figure}

Related to the surface energies are the two capillary lengths
$d_0^{\alpha}$ and $d_0^{\beta}$ defined as
\begin{equation}   
 d^{\nu}_0=\frac{\gamma_{\nu L}}
     {(\Delta c_{\nu})^2({\partial\mu}/{\partial c})} \label{caplength}
\end{equation}
which can also be expressed in terms of the Gibbs-Thomson constants
$\Gamma_{\nu}$ by
\begin{equation}
 d^{\nu}_0=\frac{\Gamma_{\nu}}{|m_{\nu}|\Delta C_{\nu}}.
\end{equation}
These are two of the physical length scales that are relevant in pattern
formation in solidification problems. In real systems, the capillary
lengths are microscopic and much smaller than all other physical length
scales in the problem. Ideally, one would like to adjust the capillary
lengths in the model to match the physical length scale ratios by
choosing suitable model parameters. Since $\partial\mu/\partial c=\lambda$, 
it follows that $d_0^\nu$ depends on $\lambda$ as
\begin{equation}   
d^{\nu}_0 \propto \frac{\gamma_{\nu L}}{\lambda}. \label{diffleng}
\end{equation}
Hence, to have small capillary lengths, one would like to increase
$\lambda$. However, $\lambda$ cannot be chosen arbitrarily large for
two reasons. First, the surface tensions themselves depend
weakly on $\lambda$ for $T\neq T_p$. As shown in Fig.~\ref{surten},
these variations amount to a few percent over the temperature 
range of interest when $\lambda$ is varied by a factor of $5$. 
Secondly, the temperature range in which equilibrium interface
solutions exist also depends on $\lambda$. 
More precisely, with a fixed value for $\lambda$, the
$\alpha$-liquid equilibrium solution does not exist if $T$ is below a
certain value, and the $\beta$-liquid interface solution
ceases to exist if $T$ is above another value, because if $T$ is too low
or too high, the free energy density loses a minimum at $\psi$ equal to
$+1$ or $-1$, respectively. We have estimated the range of temperatures in
which both solutions exist for different $\lambda$ by finding equilibrium
solutions at different temperatures. The results are shown in
Fig.~\ref{T_limit}.
We can see that this temperature range becomes
narrower when $\lambda$ increases. From now on, we fix $\lambda=2.5$
unless otherwise stated. This is a compromise between having a large
$\lambda$ and and a sufficient working temperature range in which our
two-dimensional simulations can be carried out. 

To check Young's condition, we performed two-dimensional simulations at
$T_p$ on a square grid with $\Delta x/W=0.8$. The equilibrium angles
around a trijunction were measured and found to be consistent with
Eq.~(\ref{Young}) to within a few degrees. 

For a moving interface, there are also nonequilibrium
kinetic effects related to the attachment of atoms at the interface
and solute trapping. Since we are mostly interested here in 
qualitative aspects of the growth
morphologies, we have not analyzed all these effects in
detail. We checked, however, by performing dynamical one-dimensional
simulations that nonequilibrium effects only lead to a deviation from local
equilibrium that does not exceed the Gibbs-Thomson effect caused by
interface curvature in two-dimensional simulations. 

\section{Simulations}

For our simulations, we cast the equations of motion into a
dimensionless form. For simplicity, we take $W_\phi=W_\psi=W$ and
$\tau_\phi=\tau_\psi=\tau$. By defining the dimensionless variables
\begin{equation}
\tilde{\bf r}=\frac{\bf r}{W}, \quad \tilde{t}=\frac{t}{\tau}
\end{equation}
and the new variable
\begin{equation}
\tilde{\mu}=c+A_1 h(\phi)+\frac{1}{2}A_2[1+h(\phi)]h(\psi),
\end{equation}
the equations of motion can be written in the form
\begin{eqnarray}
\frac{\partial \phi}{\partial\tilde{t}}&=&\tilde{\nabla}^2\phi
                 -\frac{\partial f}{\partial \phi}, \label{dimless1}  \\
\frac{\partial \psi}{\partial\tilde{t}}&=&\tilde{\nabla}^2\psi
                 -\frac{\partial f}{\partial \psi}, \label{dimless2}  \\
\frac{\partial c}{\partial\tilde{t}}&=&\alpha\tilde{\nabla}
    \cdot[\tilde{D}(\phi)\tilde{\nabla}\tilde{\mu}],     \label{dimless3}
\end{eqnarray}
where
\begin{equation}
\alpha=\frac{\tau D_L}{W^2}
\end{equation}
is the scaled diffusion coefficient of the impurity in the liquid, and
\begin{equation}
\tilde{D}(\phi)=(1-\phi)/2 .
\end{equation}
Instead of the concentration far from the interface, we may
also use $\eta_\beta$, the volume fraction of $\beta$ formed in 
the solid, to characterize the overall composition of the sample.
The two quantities are related by
\begin{equation}
C_\infty=(1-\eta_{\beta})C_{p\alpha}+\eta_{\beta}C_{p\beta}.
\end{equation}

In a typical directional solidification experiment, the sample is pulled
under a temperature gradient $G$ with a pulling velocity $v_p$. We
define the dimensionless temperature gradient and velocity, ${\tilde G}$
and ${\tilde v}$, by
\begin{eqnarray}
{\tilde G}&=&\frac{G}{|m_{\alpha}|\Delta C_{\alpha}} W, \\
{\tilde v}_p&=&v_p\tau/W.
\end{eqnarray}
Usually, thermal diffusion
is orders of magnitude faster than the diffusion of the impurities, and
hence we use the ``frozen temperature approximation'' which assumes
that the temperature of the system adjusts instantaneously to the
externally imposed temperature gradient. Accordingly, 
directional growth along the $x$ axis is implemented by letting
\begin{equation}
{\tilde T}={\tilde T_0}+{\tilde G}({\tilde x}-{\tilde v_p}{\tilde t}), 
\label{frozent}
\end{equation}
where ${\tilde T_0}$ is some reference temperature.

There are five different physical length scales that control the 
microstructural pattern formation: the two capillary
lengths $d_0^{\alpha}$ and $d_0^{\beta}$ defined by Eq. (\ref{caplength}), 
the two thermal lengths
\begin{equation}
l_T^{\nu} = \frac{|m_{\nu}|\Delta C_{\nu}}{G}
           =\frac{|m_{\nu}|\Delta C_{\nu}}{|m_{\alpha}|\Delta C_{\alpha}}
            \frac{W}{{\tilde G}},
\end{equation}
and the diffusion length
\begin{equation}
      l_D = \frac{D_L}{v_p}=\frac{\alpha}{{\tilde v_p}}W.
\end{equation}

Equations.~(\ref{dimless1})-(\ref{dimless3}) are integrated numerically on a
two-dimensional grid. We use $\alpha=1$, $\Delta\tilde{x}=0.8$, and
$\Delta\tilde{t}=0.1$. Zero-flux boundary conditions are applied to the
two sides that are parallel to the growth direction. There are
several features in the model that can be exploited in order to speed up
the computation. First, the phase fields $\phi$ and $\psi$ differ
significantly from $\pm 1$ only in the interfacial region, and hence we
can avoid integrating Eqs.~(\ref{dimless1}) and (\ref{dimless2}) away from
the interface. In addition, Eq.~(\ref{dimless3}) needs to be integrated
only in the liquid. Secondly, the concentration field decays exponentially
in the growth direction and varies only slowly in space in the liquid
region far ahead of the interface. Hence, we can use a coarser and coarser
grid as we move away from the interfacial region. Thirdly, in order to
simulate a semi-infinite system in the growth direction, we take advantage
of the fact that all the fields remain unchanged in the solid in the
one-sided model. Whenever the solid-liquid interface has advanced one
lattice spacing, we pull the system back by one unit and keep the
composition at the end of the liquid side at $c_{\infty}$. With all these
implementations, we are able to carry out simulations with typical lengths
in the growth direction equal to about ten times the diffusion length. 
For the results presented in this article, we chose a pulling speed such
as to have a diffusion length of $l_D=200W$. Other parameters and length
scales are listed in Table~\ref{lengscale}. 
\begin{table}
\caption{\label{lengscale} List of simulation parameters.}

\begin{center} 
\begin{tabular}{ccccc}
$\lambda$      & 2.5                    & & $l^\alpha_T/l_D$   & 0.895 \\
${\tilde G}$   & $5.5838\times 10^{-3}$ & & $l^\beta_T/l_D$    & 0.0934\\
${\tilde v_p}$ & $5\times 10^{-3}$      & & $d^{\alpha}_0/l_D$ & 
                                                $2.620\times10^{-3}$   \\
               &                        & & $d^{\beta}_0/l_D$  &
                                                $2.5119\times 10^{-2}$ \\
\end{tabular}
\end{center}
\end{table}

\section{Results}
\subsection{Dynamics of spreading}
\nobreak
Let us first concentrate on the spreading of the $\beta$ phase on the
$\alpha$ phase, starting from a single nucleus. Similarly to the 
situation considered in Trivedi's model,
the simulation is started with a homogeneous composition in the liquid and
a planar $\alpha$-liquid interface. The lateral system size $L$ is 
several times the diffusion length. Nucleations are assumed to occur
heterogeneously at the solid-liquid interface when the temperature of the
metastable interface reaches a certain undercooling with respect to the
stable solid-liquid equilibrium. The nucleation undercoolings $\Delta
T_N^{\alpha}$ ($\alpha$ on $\beta$) and $\Delta T_N^{\beta}$ ($\beta$ on
$\alpha$), shown in Fig.~\ref{schematic}, are assumed to be constant.
Accordingly, in our simulations a circular nucleus of $\beta$ phase is put
at the solid-liquid interface on one side of the box when the liquid
composition at the interface reaches the threshold for nucleation fixed by
the nucleation undercooling $\Delta T_N^{\beta}$. The radius of the
nucleus is taken to be $6W$, slightly larger than the critical radius for
nucleation. Since we are interested here in the deterministic spreading
dynamics following a single nucleation event, further nucleation
is prohibited. Multiple nucleation events will be treated in Sec. \ref{secext}.

\begin{figure}
\centerline{
 \psfig{file=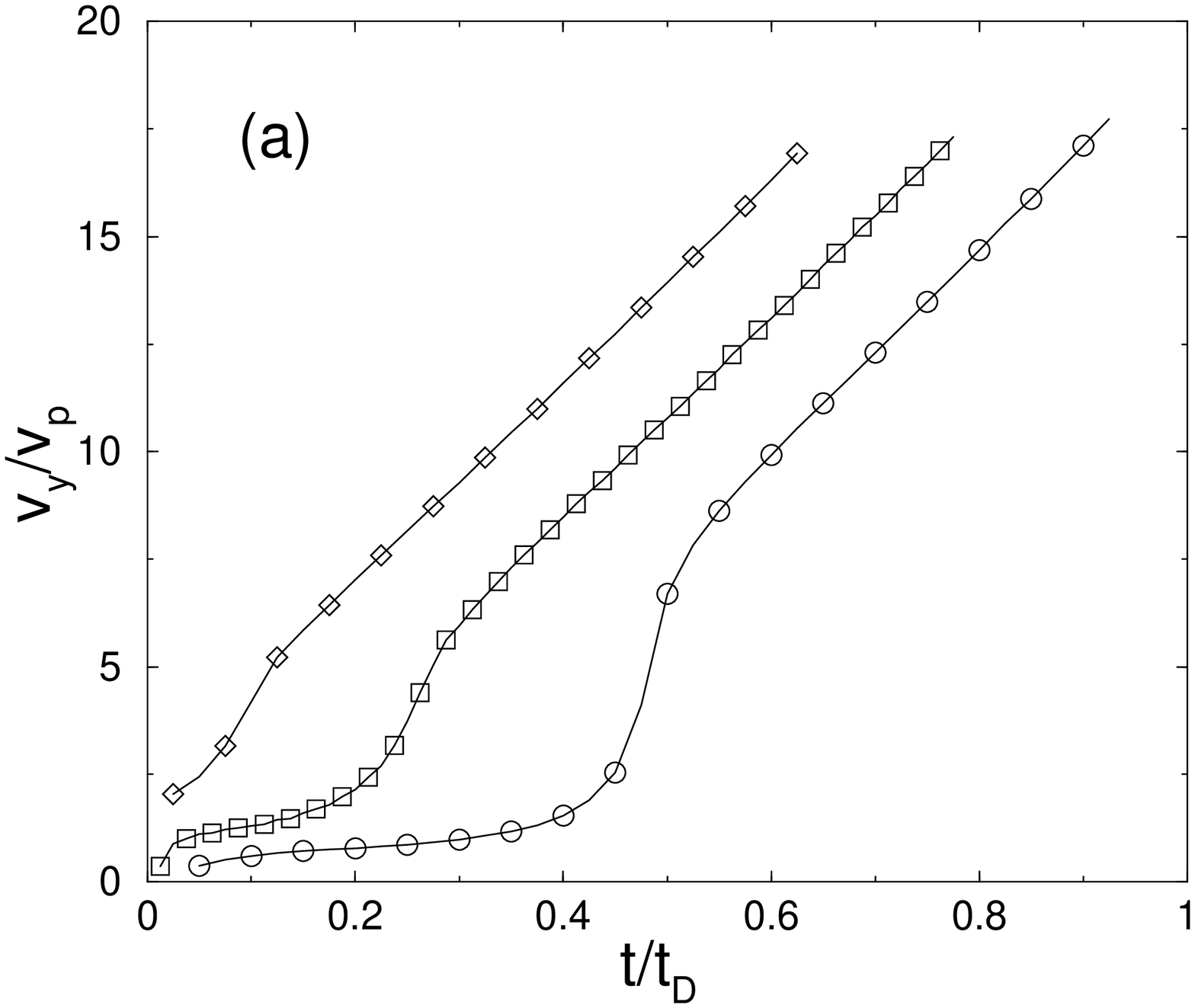,angle=-90,width=.42\textwidth}}
\centerline{
 \psfig{file=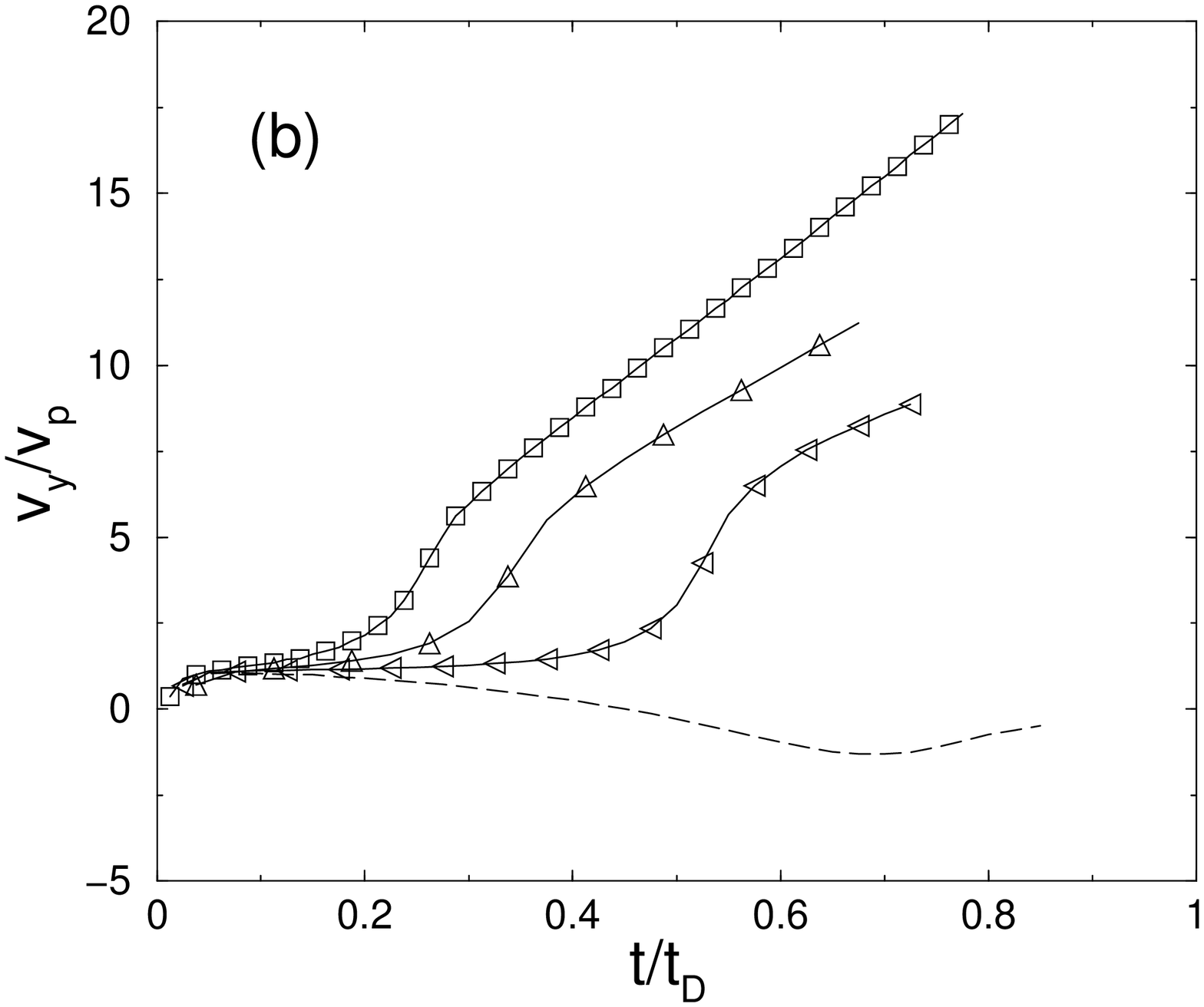,angle=-90,width=.42\textwidth}}
\centerline{
 \psfig{file=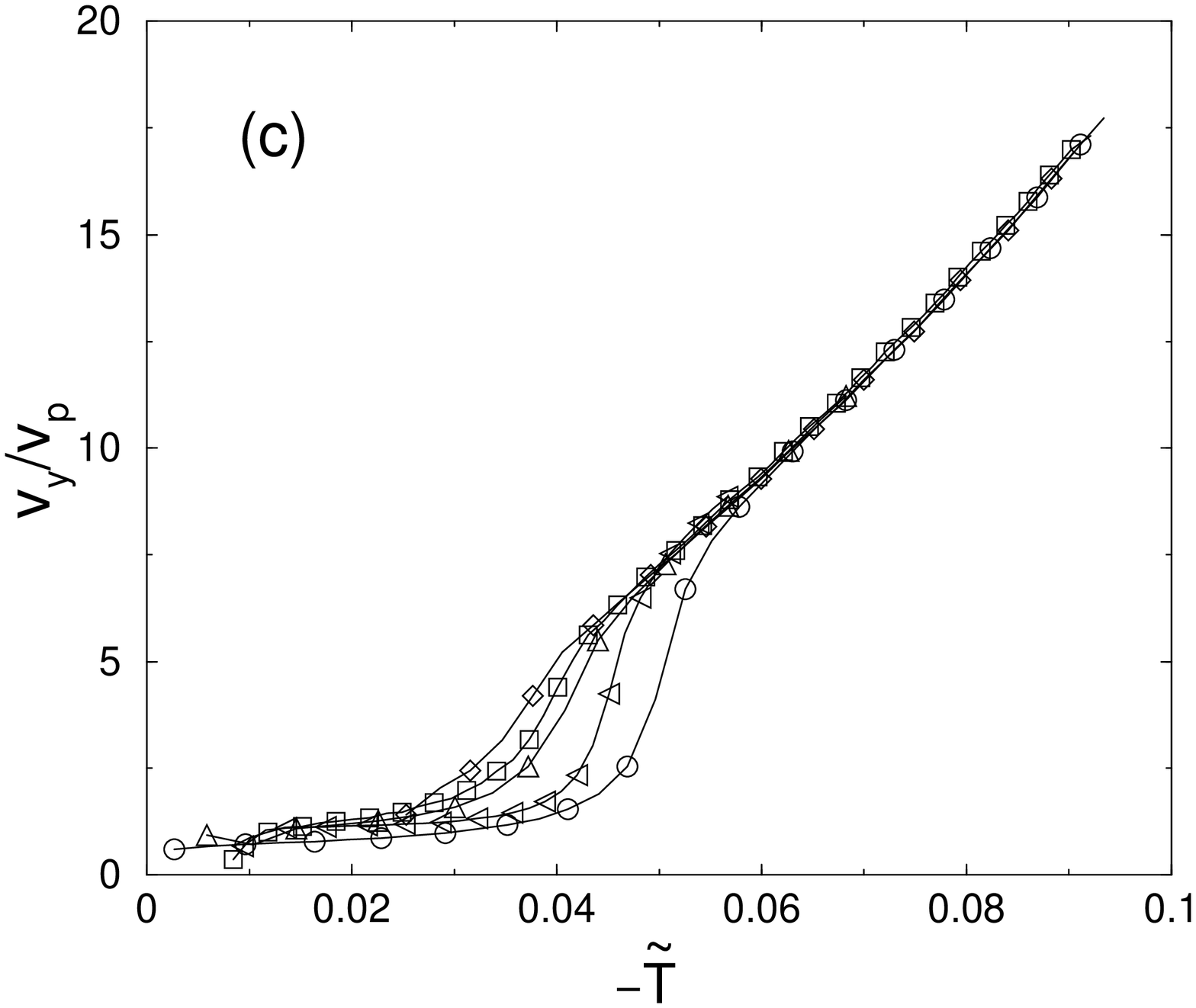,angle=-90,width=.42\textwidth}}
\caption{Plots of $v_y/v_p$ (a) versus time for constant
volume fraction and varying nucleation undercooling,
(b) versus time for constant nucleation undercooling and
varying volume fraction, and (c) versus scaled undercooling $-\tilde T$ 
of the $\alpha$L-interface with respect to $T_p$.
Circles: $\eta_{\beta}$=0.375, $\Delta{\tilde T}_N^{\beta}$=0.01878;
Squares: $\eta_{\beta}$=0.375, $\Delta{\tilde T}_N^{\beta}$=0.03129;
Diamonds: $\eta_{\beta}$=0.375, $\Delta{\tilde T}_N^{\beta}$=0.04381;
Upward triangles: $\eta_{\beta}$=0.25, $\Delta{\tilde T}_N^{\beta}$=0.03129;
Leftward triangles: 
  $\eta_{\beta}$=0.3125, $\Delta{\tilde T}_N^{\beta}$=0.03129.
The dashed line in (b) is for $\eta_{\beta}$=0.1875, 
$\Delta{\tilde T}_N^{\beta}$=0.03129, where an island is formed.
}\label{linear1}
\end{figure}

To characterize the dynamics of spreading, we recorded the position
and velocity of the trijunction point. The sideways velocity $v_y$
can be regarded as a measure of the spreading speed of the $\beta$ phase.
Figures.~\ref{linear1}(a) and \ref{linear1}(b) show plots of $v_y/v_p$ versus
time for different nucleation undercoolings and different compositions
$c_\infty$, respectively. Time is measured in terms of the diffusion time
\begin{equation}
t_D=\frac{l_D}{v_p}=\frac{D_L}{v_p^2}.
\end{equation}
Two very different regimes of spreading can be clearly distinguished.
Immediately after the nucleation, the spreading velocity 
is almost independent of the composition, but strongly depends on
the nucleation undercooling. The growth of the nucleus is
influenced only by its immediate surroundings. On the length scale of the
nucleus, which is much smaller than the diffusion length, the impurity
concentration can be considered constant and is determined only by the
nucleation undercooling. A higher $\Delta{\tilde T}_N^{\beta}$ is
equivalent to a higher supersaturation, and hence a higher growth speed. 

\begin{figure}
\centerline{
 \psfig{file=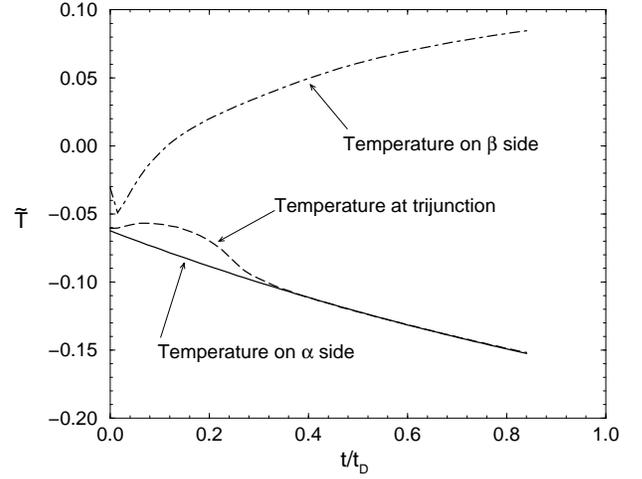,angle=-90,width=.45\textwidth}}
\caption{Scaled temperatures at the trijunction and at the solid-liquid
interface on both sides of the box. ($\Delta{\tilde 
T}_N^{\beta}$=0.03129, $\eta_\beta$=0.375.)
}               \label{undercool}
\end{figure}

At later times, the modifications of the diffusion field induced by
the growing $\beta$ phase influence the spreading dynamics, and the
spreading velocities for equal undercooling, but different $c_\infty$
start to differ [Fig.~\ref{linear1}(b)]. After a complicated transient, 
the details of which depend on the choice of parameters, $v_y$
becomes a linear function of time, which means a constant lateral
acceleration of the trijunction. This acceleration is independent of the
nucleation undercooling or the history of the system, but depends
on the composition. An explanation of this finding can be deduced 
from Fig.~\ref{undercool}, which shows the interface temperatures 
on the sides of the box and at the trijunction as functions of time. 
After the initial transient, the temperature at the
trijunction just follows the temperature on the $\alpha$ side. This
implies that the $\alpha$-liquid interface is almost planar up to the
trijunction. We can also see that during the whole time of the simulation,
the $\alpha$-liquid interface is still relaxing toward its steady state
below $T_p$. Hence, the undercooling that drives the $\beta$ phase to
spread is increasing.

In the late stages of spreading, the lateral diffusion length 
$D/v_y$ becomes much smaller than the solute boundary
layer, and is comparable to or even smaller than the tip of the
spreading finger. Therefore, the spreading speed should be
a function of local supersaturation only. To check this
assumption, we show in Fig. \ref{linear1}(c) the same velocity
curves as before, but now plotted against the undercooling
of the planar $\alpha$-liquid interface with respect to the
peritectic temperature, $-\tilde T$.
Since in our phase diagram the liquidus curves
are straight lines, this undercooling is simply proportional
to the supersaturation. The curves all collapse onto a single master 
curve after the initial transient, i.e., starting from the time
when the interface ahead of the trijunction has become flat.
This master curve is not linear, and does not smoothly 
extrapolate to zero. We did not attempt to calculate it
theoretically. We expect that its detailed form should depend 
on the characteristics of the trijunction, and in particular 
on the angles between the different interfaces.
More theoretical and numerical work would be needed to
elucidate in detail the role of the various material parameters.
Remarkably, similar observations were very recently reported
in experiments on a transparent organic eutectic alloy \cite{FaiAka}
during spreading of the secondary phase on a planar interface
of primary phase. The spreading speed of the secondary phase
showed an approximately linear increase with time, and the
data could also reasonably well be rescaled onto an analogous 
master curve.

Since the $\alpha$-liquid interface far ahead of the trijunction
stays fairly planar before the arrival of the $\beta$ phase, 
the time dependence of the temperature on the $\alpha$
side can be well described by the Warren-Langer approximation
\cite{WarrenLanger}. The rate of change of the supersaturation
is solely determined by the composition $c_\infty$, which explains why 
the final slope of the curves in Figs.~\ref{linear1}(a) and \ref{linear1}(b)
depends on $c_\infty$ but not on the nucleation undercooling. 

A completely different behavior is observed when the composition is
sufficiently low. As shown in Fig.~\ref{linear1}(b) (dashed line), the
initial spreading speed is the same as for the other runs. However, at
later times, the spreading slows down and the trijunction point turns
around such that $v_y$ becomes negative. Instead of a band, an
isolated island of $\beta$ phase is formed. This phenomenon will be 
addressed in detail in Sec. \ref{islandsec} below.

In the final regime of spreading, when the lateral speed becomes much
larger than the pulling speed, the lateral diffusion length $D_L/v_y$
becomes comparable to the radii of curvature close to the trijunction
point. In free growth, such conditions are reached only at very large
solidification speeds. Under these circumstances, it is clear that the
phase-field model does no longer reflect quantitatively the sharp-interface 
equations, since it contains corrective terms due to the finite 
width of the interface. For instance, we observed a violation of Young's
condition at the trijunction point. More precisely, the angles between the
interfaces, obtained by taking the tangent vectors to the $\phi=0$ and
$\psi=0$ level curves at the trijunction, are still consistent with local
equilibrium, but the solid-solid interface is highly curved on a length
scale comparable to the width $W$ of the diffuse interface. This is due to
the fact that the diffusivity varies smoothly within the diffuse interface,
and hence the part of the solid-solid interface near to the trijunction is
still able to move. As a result, the angles between the interfaces, seen
on a macroscopic scale, differ from the local equilibrium angles.
For the purpose of the present study, where we are mainly interested 
in the qualitative features of the microstructures, we did not 
investigate this effect quantitatively. Let us remark, however, that
such effects may not be simply an artifact of the phase-field model,
but may have a physical significance for high growth speeds if the
relaxation of the trijunction toward local equilibrium occurs on a
time scale comparable to the time of diffusion through the trijunction 
region.

\subsection{Morphology transition}
\label{islandsec}
\begin{figure}
\centerline{
 \psfig{file=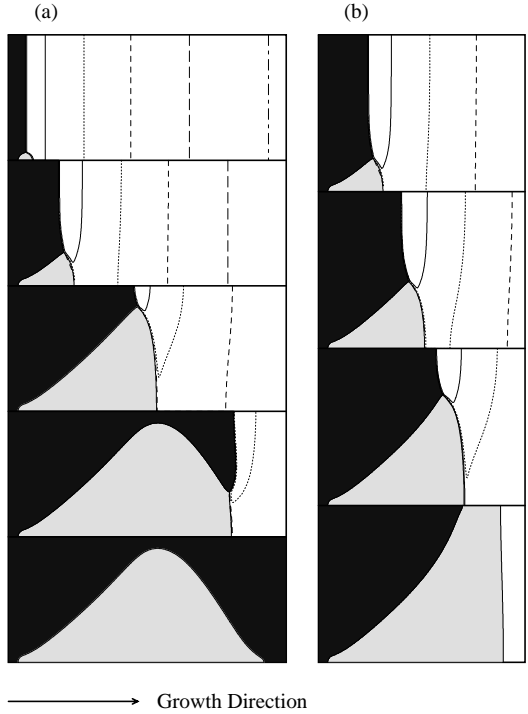,angle=-180,width=.45\textwidth}}
\caption{Spreading of a single $\beta$ nucleus over the $\alpha$
phase with $\Delta{\tilde T}_N^{\beta}=0.02504, \eta_{\beta}=0.3125$.
(a) Island formation at $L/l_D=0.512$, (b) Band formation at $L/l_D=0.64$.
Time increases from top to bottom. (Dark region, $\alpha$ phase; light 
region, $\beta$ phase; unshaded region, liquid. The isoconcentration lines in
the liquid are evenly spaced in $c$.)}
\label{betaspread}
\end{figure}

\nobreak
The results of the preceding section were obtained for systems with
lateral extensions of several times the diffusion length. For some sets of
parameters, a surprising event occurs when the system size is reduced
while all other parameters are kept constant. After some time, the
spreading slows down, and the trijunction point may even turn around, such
that $v_y$ becomes negative. As a result, the trijunction travels back to
the wall where it originated, and an isolated island of $\beta$ phase, or
partial band, is formed. It hence appears that complete spreading is
easier to achieve in larger systems, a quite counterintuitive result.
Figsures~\ref{betaspread}(a) and \ref{betaspread}(b) show time series of 
typical snapshot pictures for the formation of an island and a band, respectively. 
The scales are the same on both axes and in both figures. Isoconcentration
lines in the liquid are also shown. It can be seen that a lateral
concentration gradient builds up in the liquid. This concentration
gradient plays an important role in the interpretation of the
morphological transition from islands to bands and will be discussed
below.

To study more systematically the conditions for the formation of 
islands, we performed simulations with various lateral system 
sizes $L$ and compositions $c_\infty$, with the following results1 
\begin{enumerate}
\item At a fixed $\Delta T_N^{\beta}$, there exists a critical composition
$c^*$ such that if $c_\infty<c^*$ the $\beta$ phase always forms islands.
This critical composition decreases as the nucleation undercooling increases. 

\item At a fixed $\Delta T_N^{\beta}$ and if $c_\infty>c^*$, there exists
a critical lateral system size $L_c(c_\infty)$ such that if the lateral 
system size $L>L_c$ the $\beta$ phase spreads completely and forms bands, 
whereas for $L<L_c$ it forms islands. $L_c$ decreases when either $c_\infty$ 
or $\Delta T_N^{\beta}$ increases. 

\item When $c_\infty<c^*$, such that the $\beta$ phase
always forms islands, the final shape (and also the size) of
the islands is independent of $L$ when $L$ is larger than a certain size. 
\end{enumerate}

\begin{figure}
\centerline{
 \psfig{file=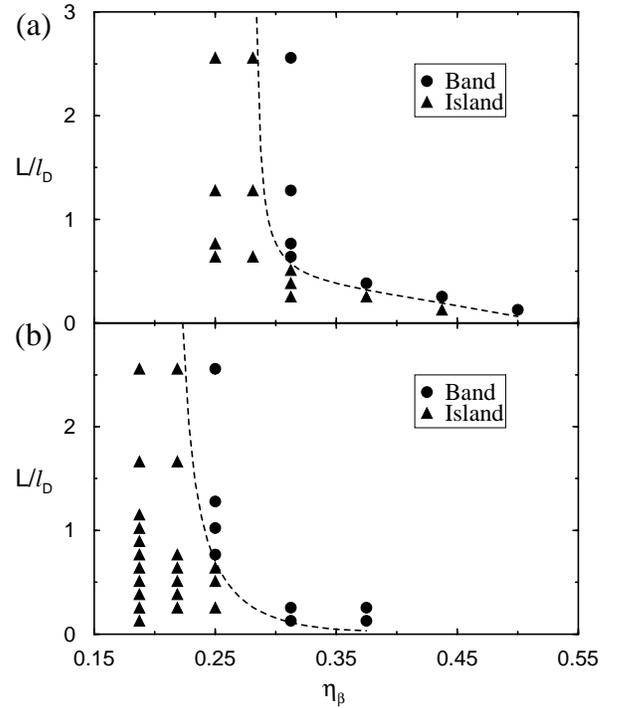,angle=-180,width=.45\textwidth}}
\caption{Morphology map at (a) $\Delta{\tilde T}_N^{\beta}=0.02504$
and (b) $\Delta{\tilde T}_N^{\beta}=0.03129$.}
\label{morpho.ps}
\end{figure}

Figure~\ref{morpho.ps} shows the final morphology of the system for
different $c_\infty$ (or equivalently different $\eta_{\beta}$) and
for different nucleation undercoolings. The dashed lines in
Figs.~\ref{morpho.ps}(a) and (b) represent an estimate for the critical
system size $L_c(c_\infty)$ for the transition from bands to islands. 
It can be seen that both $L_c$ and $c^*$ are smaller for higher 
$\Delta{\tilde T}_N^{\beta}$. 

The existence of the critical size $L_c$ can be understood by noticing
that both $\alpha$ and $\beta$ phases have to reject impurities in order
to grow, but the concentration jump at the $\alpha$-liquid interface is
larger than that at the $\beta$-liquid interface. Since $\beta$ is the stable
phase below $T_p$, there exists a driving force for the $\beta$ phase to
spread. On the other hand, as $\beta$ rejects fewer impurities than
$\alpha$, the impurity concentration in front of the $\beta$ phase
rapidly decreases after the nucleation. This creates 
a lateral concentration gradient and hence
an impurity flow from the $\alpha$ to the $\beta$ side as can be clearly
seen in Fig.~\ref{betaspread}. This lateral impurity backflow will
accelerate the growth of $\alpha$ and slow down the growth of $\beta$ and
hence there is a competition between the two phases. 

To be more precise, we can consider the following scaling argument.
Let us assume for simplicity a constant spreading speed $v_s$ for 
the $\beta$ phase. 
Then the time required by the $\beta$ phase to spread across the system
is $L/v_s$. On the other hand, impurities diffuse laterally through 
the system on a time scale of $L^2/D_L$. If $L/v_s<L^2/D_L$, 
the $\beta$ phase is able to spread over the $\alpha$
phase before a significant impurity backflow can occur. If the
opposite is true, the impurities have enough time to diffuse 
and the growth of the $\beta$ phase is slowed down. Hence, the
critical system size is given by
\begin{equation}
L_c\sim \frac{D_L}{v_s}. \label{Lc}
\end{equation} 
Another way to interpret the above criterion is to note that the
``diffusion speed'', which is roughly the speed of the impurities diffusing
laterally through the system, is given by $D_L/L$. If the diffusion speed
is smaller than $v_s$, spreading occurs, and Eq.~(\ref{Lc}) follows
immediately.

Clearly, the above argument is only qualitative. We have assumed 
a constant lateral spreading speed in Eq.~(\ref{Lc}), although
Fig.~\ref{linear1} shows that the spreading speed varies
with time, and hence we can give no explicit expression for $v_s$
as a function of composition and nucleation undercooling. However,
we can see from Fig.~\ref{linear1} that, for any given time, the
instantaneous spreading speed increases with increasing nucleation
undercooling and increasing volume fraction of $\beta$ phase.
This observation, together with Eq. (\ref{Lc}), allows us to
understand qualitatively the shape of the curves $L_c(c_\infty)$
in Fig.~\ref{morpho.ps}. In addition, this criterion allows
understanding of the striking finding that spreading is easier in
large systems that in small ones.

\subsection{Banding and island formation}
\nobreak
So far, we have concentrated on how a single $\beta$ nucleus 
spreads on the $\alpha$ phase. It is natural to ask what happens 
if renucleation is allowed. This is a complicated problem
since nucleation is an inherently stochastic phenomenon, which 
cannot be consistently treated within our deterministic model. 
However, we can try to gain some insight by incorporating
nucleation phenomenologically. We will proceed in two steps.
First, we treat repeated nucleation in small samples by deterministic 
rules to make contact with the recent experiments 
in the Sn-Cd alloy system \cite{Park98}. Then, in the next section, 
we investigate the influence of multiple stochastic nucleations
on the pattern formation dynamics in large systems.

\begin{figure}
\centerline{
 \psfig{file=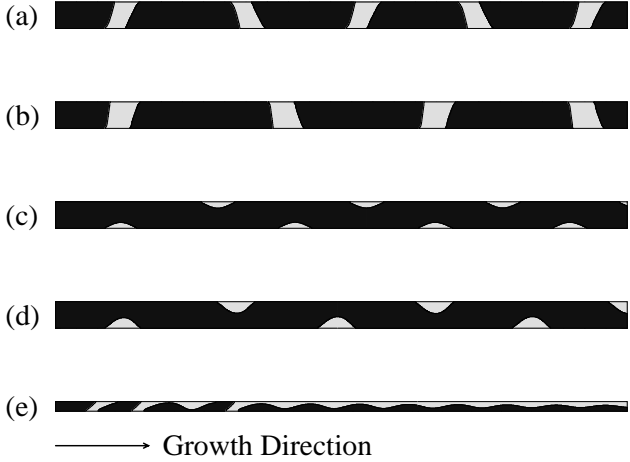,angle=-90,width=.5\textwidth}}
\caption{Microstructures obtained from simulation for 
(a) $\Delta{\tilde T}_N^{\alpha}$=0.13565,
$\Delta{\tilde T}_N^{\beta}$=0.04381, $L/l_D$=0.512, $\eta_{\beta}$=0.125,
(b) $\Delta{\tilde T}_N^{\alpha}$=0.17440,
$\Delta{\tilde T}_N^{\beta}$=0.05633, $L/l_D$=0.512, $\eta_{\beta}$=0.125,
(c) $\Delta{\tilde T}_N^{\beta}$=0.02504, $L/l_D$=0.512, $\eta_{\beta}=$0.05,
(d) $\Delta{\tilde T}_N^{\beta}$=0.03129, $L/l_D$=0.512, $\eta_{\beta}$=0.075
and (e) $\Delta{\tilde T}_N^{\alpha}$=0.01938,
$\Delta{\tilde T}_N^{\beta}$=0.03129, $L/l_D$=0.128, $\eta_{\beta}$=0.4375.
(Dark region, $\alpha$ phase; light region, $\beta$ phase.)}
\label{tubes}
\end{figure}

For solidification in small systems, it can be assumed that
nucleation occurs predominantly at the container walls close
to the solid-liquid interface. The density of nuclei
and the nucleation rate are very rapidly varying functions of the
composition. Therefore, it seems reasonable to assume that a nucleus will
form as soon as the concentration in the liquid exceeds the threshold
corresponding to the nucleation undercooling. Accordingly, we
incorporate repeated nucleation by the following rules
(the nucleation of $\alpha$ is handled like the nucleation
of $\beta$ before, by placing a small circular nucleus
at the solid-liquid interface).
\begin{enumerate}
\item A nucleus of the new phase is placed at one side of the
container as soon as the undercooling of the interface exceeds
the nucleation undercooling. If both sides of the box reach the
threshold at the same time, one side is chosen at random.
\item Once a germ has nucleated, further nucleation of the same
phase is prohibited
until the germ has either completely spread across the system
or completed the formation of an island.
\end{enumerate}
We consider now the two nucleation undercoolings as free parameters,
and study different cases. When both $\Delta T_N^{\alpha}$ and $\Delta
T_N^{\beta}$ are large enough such that the newly nucleated phase spreads
completely before the original phase is able to renucleate again,
banded structures are obtained. For a smaller $\Delta T_N^{\beta}$, the 
$\beta$ phase does not spread completely, but forms an island.
The $\alpha$ phase overtakes the $\beta$ phase 
and continues to grow until the nucleation threshold for $\beta$ is
reached again. The islands of $\beta$ phase form alternately on
each side. This is a result of the history of the system: as a result of
the formation of the previous island, the concentration of impurities
is lower on the side where the last island occurred, and hence nucleation
of $\beta$ is favored at the other side. Examples of these banded and
island structures are shown in Fig.~\ref{tubes}(a) to Fig.~\ref{tubes}(d). 
The scales on both axes in these figures are the same, but
Fig.~\ref{tubes}(e) has a different scale from Figs.~\ref{tubes}(a)--(d).
This last picture was obtained by a simulation at much smaller 
$\Delta T_N^{\alpha}$, and the lateral size of the system is
smaller. In this case, an oscillatory structure is 
obtained which tends to approach a coupled growth steady
state after a complicated transient.

These results are in good qualitative agreement with microstructures
obtained in small samples of Sn-Cd alloy \cite{Park98}. In the
experiments, islands tend to form always on the same side of the sample.
We believe that this is due to a slight lateral temperature gradient
across the sample, which is always present in experiments.

\subsection{Nucleation controlled microstructures in spatially extended
systems}
\label{secext}
\nobreak
Until now, we have mainly focused on the microstructures formed in small
samples. It is interesting to ask what kinds of structure are to be expected
in large samples in the absence of convection. This situation could be
achieved either in quasi-two-dimensional thin samples, or in a
microgravity environment. To model the microstructure formation, larger scale
computations were carried out. However, in a spatially extended system, 
multiple nucleations are unavoidable and must be incorporated in a 
way that is consistent with the predictions of classical nucleation theory. 
We chose to extend our model by incorporating the effects of multiple 
nucleation in a phenomenological manner.

\begin{figure}
\centerline{
 \psfig{file=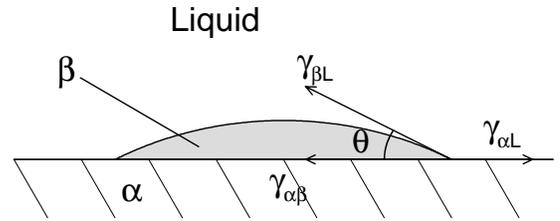,angle=-90,width=.4\textwidth}}
\smallskip
\caption{Sketch of a critical nucleus in the spherical cap model for
heterogeneous nucleation.}                   \label{droplet}
\end{figure}

For the nucleation of the $\beta$ phase on a planar $\alpha$ front (the
same arguments also apply to the nucleation of $\alpha$ on $\beta$),
classical nucleation theory predicts the nucleation rate
\begin{equation}
  I=I_0~e^{-\Delta F^*/k_B T},         \label{nuc1}
\end{equation}
where $I_0$ is a constant prefactor (with dimension equal to the number
of nucleations per unit volume per unit time) and $\Delta F^*$ is the
activation energy for heterogeneous nucleation. Assuming that the
critical nucleus is a spherical cap on a planar substrate (the spherical
cap model), $\Delta F^*$ is given, respectively, in two and three
dimensions by
\begin{equation}
\Delta F^*=\left\{
            \begin{array}{ll}
          \frac{\gamma_{\beta L}^2}{\Delta F_B}
            \times\frac{\theta^2}{\theta-(1/2)\sin 2\theta}, & {\rm 2D} \\
          \frac{\gamma_{\beta L}^3}{\Delta F_B^2}\times
            \frac{16\pi (2+\cos\theta)(1-\cos\theta)^2}{12}, & {\rm 3D},
            \end{array}
           \right .
\end{equation}
where $\Delta F_B$ is the difference between the bulk free energy of the
$\beta$ phase and of the liquid phase. The contact angle $\theta$, as
shown schematically in Fig.~\ref{droplet}, is determined by the balance of
surface tensions parallel to the substrate,
\begin{equation}
\gamma_{\alpha L}=\gamma_{\alpha\beta}+\gamma_{\beta L}\cos\theta.
                                                         \label{nuc3}
\end{equation}
Assuming that the system is locally in thermodynamic equilibrium, it can be
shown that $\Delta F_B$ is proportional to $(T-T_p)$
\cite{Christian65}, such that for a quasi-two-dimensional system the
nucleation rate for $\beta$ on $\alpha$ can be written as
\begin{equation}
 I=\left\{
      \begin{array}{ll}
         I_{2D}\exp[-A/(T-T_p)^2] \label{nuclrate} & {\rm if}\quad T<T_p \\
         0  & {\rm if}\quad T\geq T_p
            \end{array}
           \right .
\end{equation}
where $A$ is a constant, and $I_{2D}$ has now the dimension of 
number of nucleations per unit time and per unit length of the 
interface. A similar expression
with $w=0$ when $T\le T_p$ holds for $\alpha$ nucleating on
$\beta$. The 3D form of $\Delta F^*$ is used in deriving
Eq.~(\ref{nuclrate}) since, in practice, the size of a nucleus is still
much smaller than the thickness of a thin sample. Equation~(\ref{nuclrate})
determines the local nucleation rate and hence the probability per unit
time of a nucleus forming as a function of the local temperature at the
solid-liquid interface.

Unfortunately, both experimental and theoretical estimates of
the free energy barrier and the kinetic prefactor are scarce
in the context of heterogeneous nucleation, since the actual
values may depend on complicated details of the interfacial
structure. Since, in the present study, we focus on morphological
aspects of the large scale structure, we decided to treat the
two quantities as free parameters. Moreover, we want to compare
the stochastic simulations to the deterministic runs of the
preceding sections. Consequently, we may eliminate one of 
those two parameters by the requirement of recovering the
rules used previously. That is, in the deterministic simulations
a nucleus was put at the solid-liquid interface when it reached
the predetermined nucleation undercooling. In the stochastic
runs, nucleation should therefore occur with probability $1$ for
the same interface temperature. This condition will lead to a
relation between the prefactor and the energy barrier in the
nucleation rate.

To proceed, let us first specify how we treat nucleation in the
simulation algorithm. The interface is scanned at a regular time
interval $\Delta t_N$, and nucleation is attempted at points 
regularly spaced by a distance $\Delta s_N$ along the interface. 
The nucleation rate may be rewritten as 
\begin{equation}
I={w(T)\over \Delta t_N \Delta s_N},
\end{equation}
where
\begin{equation}
 w=\left\{
      \begin{array}{ll}
         w_0 \exp[-A/(T-T_p)^2] \label{nuclratew} & {\rm if}\quad T<T_p \\
         0  & {\rm if}\quad T\geq T_p
            \end{array}
           \right.
\end{equation}
is a dimensionless function of the interface temperature.
At each test point, a nucleus is generated with probability $1$
if $w>1$, and with probability $w$ otherwise. That is, if
$w<1$, a random number $\xi$ uniformly distributed between $0$
and $1$ is drawn, and a nucleus is generated if $\xi<w$. As
before, the nucleus is spherical and has a size of $6W$.
A possible drawback of the procedure outlined above is that
the actual nucleation rate depends on the values chosen for
$\Delta t_N$ and $\Delta s_N$. However, it is reasonable to
assume that the microstructures should not depend too sensitively
on the choice of these parameters as long as their
values are much smaller than the time and length scales
of the pattern formation process.

\begin{figure}
\centerline{
  \psfig{file=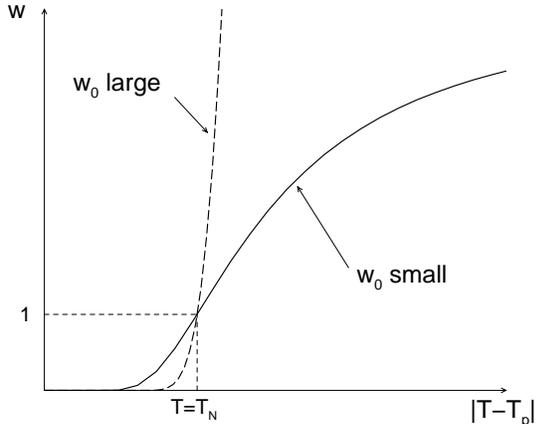,angle=-90,width=.4\textwidth}}
\caption{Schematic plot of the dimensionless 
nucleation rate $w(T)$ versus temperature for
two different choices of $w_0$ and $A$.}                   
\label{nurate}
\end{figure}

Now we can relate the prefactor and the barrier in the
nucleation rate. In the preceding sections, nuclei were 
introduced deterministically when the nucleation undercooling 
was reached, that is, at a nucleation temperature
\begin{equation}
T_N^\beta = T_p - {\Delta T_N^\beta \over 1-{m_\beta / m_\alpha}}.
\end{equation}
This implies that in the stochastic algorithm we must choose
\begin{equation}
w(T_N^\beta) = w_0 \exp[-A/(T_N^\beta-T_p)^2] = 1.
\label{nuconstraint}
\end{equation}
Given this condition, there is only one
remaining free parameter; we choose the dimensionless
kinetic prefactor $w_0$. This parameter controls the
temperature range over which nucleations occur.
Indeed, $w(T)$ is a function that rapidly increases in
a narrow temperature range around $T_N^\beta$. Increasing
$w_0$ and $A$ simultaneously while respecting the constraint
Eq. (\ref{nuconstraint}), the rise of the nucleation
rates becomes sharper, as shown schematically in Fig.~\ref{nurate}.
Now consider an $\alpha$-liquid interface during its transient,
when the temperature at the interface is decreasing
from above $T_p$ to its steady-state temperature. If the temperature
range over which the nucleation rate increases significantly is narrow,
all nucleation events will occur almost at the same time when the
interfacial temperature coincides with $T_N^\beta$. As a result, the
mean separation between the nuclei will be small. On the other hand, if
$w_0$ is small, nuclei appear with a broader spread in interfacial 
temperature, and the mean separation between the nuclei will be larger. Since
the mean separation between the nuclei plays a similar role as the system
size in a small sample experiment, we might expect a morphology
transition from bands to islands as $w_0$ is increased.

\begin{figure}
\centerline{
 \psfig{file=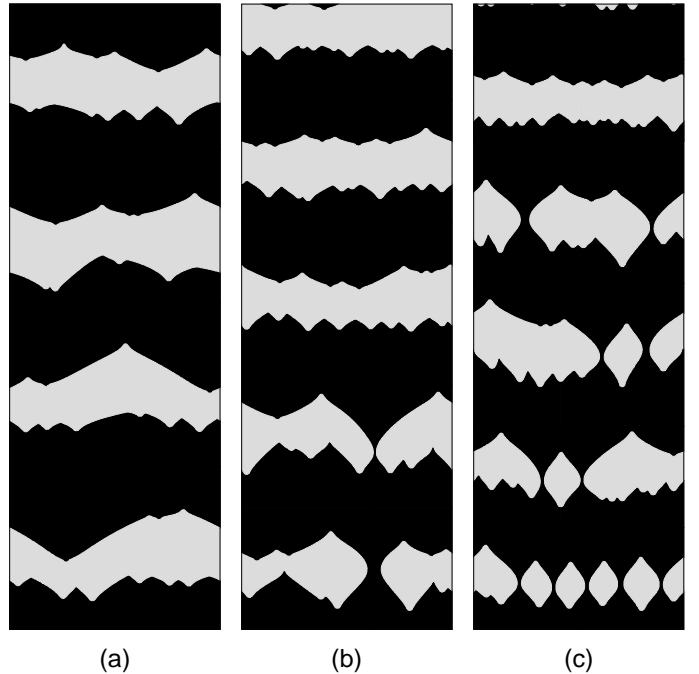,angle=0,width=.5\textwidth}}
\caption{Microstructures obtained from simulations with stochastic
nucleation events for $L/l_D=2.048$, $\eta_{\beta}=0.25$,
$\Delta{\tilde T}_N^{\alpha}=\Delta{\tilde T}_N^{\beta}=0.03129$, and 
(a) $w_0=5\times10^{3}$, (b) $w_0=5\times10^{4}$, (c) $w_0=5\times10^{5}$.  
Growth is from bottom to top; dark regions, $\alpha$ phase; light 
regions, $\beta$ phase.}
\label{stochastic1}
\end{figure}
Figure~\ref{stochastic1} shows the microstructures obtained in simulations 
for small $w_0$, ranging from $5\times10^{3}$ to $5\times10^{5}$. 
The lateral system size is about twice the diffusion length, and 
we use periodic boundary conditions in the direction perpendicular 
to the temperature gradient. The lateral spreading of multiple nuclei 
leads to a jagged morphology. Each V-shaped site in the figures 
indicates a nucleation event (either $\beta$ on $\alpha$ or
$\alpha$ on $\beta$). A transition from irregular banded
structures to islands can be observed as $w_0$ increases.
Note, however, that nucleation events occur in bursts, leading
to a spatial periodicity along the growth direction that can
be clearly distinguished in Fig.~\ref{stochastic1}(c). This
means that there is still a ``banding cycle,'' now consisting
of layers of a two-phase composite structure ({\em particulate}
structure) and layers of pure $\alpha$ matrix.

\begin{figure}
\centerline{
 \psfig{file=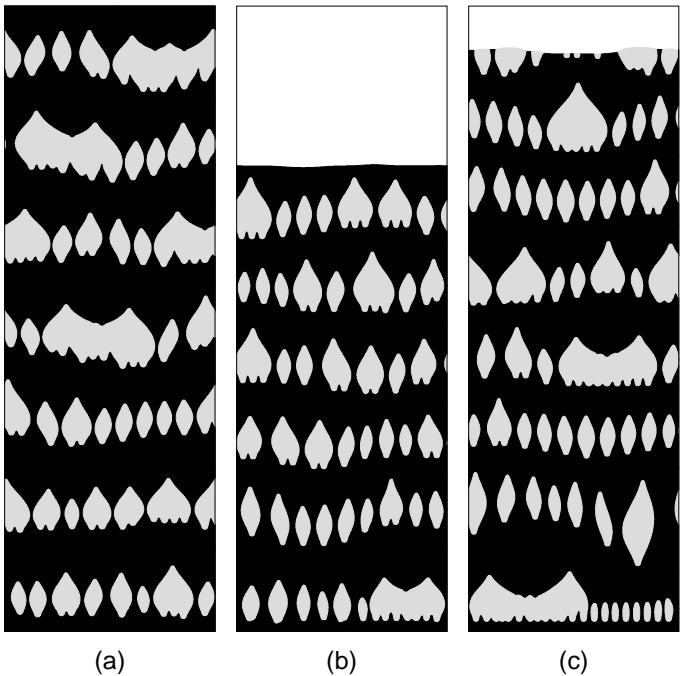,angle=0,width=.5\textwidth}}
\caption{Microstructures obtained from simulations with stochastic
nucleation events: (a) $w_0=5\times10^{11}$, (b) $w_0=5\times10^{26}$,
(c) $w_0=5\times10^{43}$; other parameters as in 
Fig.~\protect\ref{stochastic1}. The unshaded region is the
liquid.}                        
\label{stochastic2}
\end{figure}
The values for $w_0$ in Fig.~\ref{stochastic1} are somewhat
small. The attempt rate $w_0$ should be related to the rate $I_0$ in
Eq.~(\ref{nuc1}), which is typically about $10^{30}$ nuclei/${\rm
cm}^3\,{\rm s}$ for heterogeneous nucleation in metallic systems
\cite{Trivedi95}. Hence we investigated the microstructures formed
with larger values of $w_0$, ranging from $5\times10^{11}$ to 
$5\times10^{43}$. In this range, we always obtain island structures 
that look qualitatively similar (Fig.~\ref{stochastic2}), and
not too different from Fig.~\ref{stochastic1}(c). This implies that the
microstructures obtained with multiple nucleation events are not very
sensitive to $w_0$ when $w_0$ is larger than some critical value.
Hence we would expect predominantly island structures in spatially
extended systems.

The last statement, however, is valid only for the quite restrictive
assumptions made in our model. Most importantly, we have assumed
that the probability of nucleation depends only on the composition
in the liquid, and not on the local geometry of the interface.
This neglects the presence of grain boundaries and impurities
(bubbles, inclusions) which can considerably enhance nucleation.
Such heterogeneities broaden the distribution of the nucleation
rate as a function of temperature, and would hence favor bands.
The evolution of the grain structure could in principle be modeled
by including the local crystalline orientation as an additional
order parameter. An interesting perspective is that the interplay
between nucleation at grain boundaries and spreading might select
a certain grain size, since for small grains a spreading phase
can engulf and hence ``heal'' grain boundaries, whereas for large 
grains nucleation at the solid-liquid interface (as modeled 
in our simulations) may occur and lead 
to the formation of new grains. Such a study, however,
is largely beyond the scope of this article.

\section{Summary and Conclusions}

We have developed a phase-field model to
investigate a class of banded microstructures 
that form during the directional solidification of peritectic alloys under purely
diffusive growth conditions.
We focused on a regime of large thermal gradients and low pulling speeds where both phases
are morphologically stable and the interface dynamics is controlled by
a subtle interplay between the growth and nucleation of two competing solid phases,
rather than by the morphological instability of one phase. 
We restricted our attention to 
a generic peritectic phase diagram that simplified both the models and the computations,
but our approach is in principle flexible enough to be extended to phase 
diagrams of specific materials. 

The two-dimensional simulations of this model have shed light on three main aspects
of banding: the transition from islands to bands that has been observed in
narrow samples where convection is suppressed \cite{Park98}, the associated dynamical
spreading of one phase onto the other, 
and the type of structures that one would expect to form in wide samples
under purely diffusive growth conditions that are not presently accessible
in an earth-based laboratory, at least for the alloys investigated to date.

We have shown that the transition from islands to bands
can be understood in terms of a competition
between the lateral spreading of $\beta$ on $\alpha$  
and the diffusive backflow of rejected impurities from $\alpha$ to $\beta$.
This competition leads to the surprising result that bands tend to form more
easily in wider samples, in qualitative agreement with recent experiments in the
Sn-Cd alloy carried out in small samples in order to suppress convection \cite{Park98}.
The critical system size $L_c$ at which this transition occurs depends on the nucleation
undercooling for the $\beta$ phase that influences the spreading rate, 
and on the alloy composition, with $L_c$ becoming infinite  
when the volume fraction of the $\beta$ phase falls below a minimum value
necessary for band formation. The
influence of other parameters, in particular the form of the phase diagram
and the equilibrium angles at the trijunction, has not been 
investigated in detail here. 

When the peritectic phase fully covers the parent phase,
its spreading dynamics is characterized by a remarkably uniform 
acceleration of the moving trijunction that depends on the composition,
but not on the nucleation undercooling. This acceleration originates
from an increase with time of the local supersaturation  
(driving force for spreading) associated with the relaxation
of the planar parent phase ahead of the trijunction to its 
steady state below $T_p$, together with a direct relationship 
between the instantaneous speed of the trijunction and this driving 
force that depends on the material properties, but not on 
the history or the overall composition of the sample.
Moreover, the relative angles between phase boundaries at the trijunction 
during rapid spreading depart significantly from those prescribed by 
Young's condition, indicating a strong departure from local equilibrium. 
Both predictions might be experimentally testable in transparent 
organic eutectic systems that exhibit similar spreading
transients before coupled growth is established \cite{FaiAka}.

The formation of multiple nuclei in wide samples ($L\gg L_c$)
adds a stochastic element to the interface dynamics that renders 
the range of possible patterns even richer. One can nonetheless 
distinguish two basic types of structure that can be understood 
within the framework of the single island to band transition
in narrow samples ($L\sim L_c$). The first is a discrete banded 
structure made up of separate jagged bands
that span the whole width of the sample. The second is 
a {\it particulate} banded structure made up of
approximate rows of particles (islands) of the peritectic phase embedded
in the matrix of the parent phase. The banded (particulate)
structure is naturally selected if the mean distance between
nuclei is larger (smaller) than the critical sample width $L_c$ 
for the island-band transition, and, moreover, simulations reveal that
the particulate structure is preferred if nucleation 
is assumed to follow a classical nucleation law. Even though we modeled  
patterns in wide samples with such a law, we expect the transition from
a discrete to a particulate banded structure to be generally governed by the
mean distance between nuclei even if other nucleation mechanisms 
(such as wall-induced nucleation and nucleations at grain 
boundaries) play a dominant role. Both types of structure could 
conceivably coexist in the same sample if nucleation
conditions change during growth.

There are a number of possible extensions of the present study.
One is to investigate the patterns that form for a somewhat larger range of pulling
speeds where the parent phase is morphologically unstable, but the peritectic
phase is still linearly stable. Another is to incorporate the influence of convection in
a fully consistent way to make contact with experiments over a wider
range of sample sizes, which is now possible
within a phase-field context \cite{Becetal,Tonetal}.  

\acknowledgments
This research has been supported by the NASA Microgravity Research Program
under Grant No. NAG8-1254 and by U.S. DOE grant No. DE-FG02-92ER45471. 
Computations were made possible by an allocation of time at
the Northeastern University Advanced Scientific Computation Center (NUASCC).
We thank J.~S.~Park, R.~Trivedi, S.~Akamatsu, and G.~Faivre for many
helpful discussions and for giving us access to their experimental results
prior to publication. 

\appendix
\section{Results of the common tangent construction}

The common tangent construction allows one to determine the
equilibrium composition for two-phase equilibrium for given
bulk free energies $f_\nu$ of the two phases. For two-phase 
equilibrium, the bulk phases must have equal chemical potentials
$\mu=df_\nu/dc$ and grand potentials $\Omega=f_\nu -\mu c$.
Solving the resulting equations for our model bulk free energies,
we find for solid-liquid equilibrium
\begin{eqnarray}
c^{\alpha,\beta}_L&=&\frac{(B_1\pm\frac{1}{2}B_2)}{(A_1\pm\frac{1}{2}A_2)}
                        +A_1, \\
c^{\alpha,\beta}_S&=&\frac{(B_1\pm\frac{1}{2}B_2)}{(A_1\pm\frac{1}{2}A_2)}-A_1
                        \mp A_2.
\end{eqnarray}
The upper (lower) sign is for the $\alpha$ ($\beta$) phase.
\begin{flushleft}
For solid-solid equilibrium, we get
\end{flushleft}
\begin{eqnarray}
c^{\alpha}_{SS}&=&\frac{B_2}{A_2}-A_1-A_2,  \label{SS1}  \\
c^{\beta}_{SS}&=&\frac{B_2}{A_2}-A_1+A_2.   \label{SS2}
\end{eqnarray}
For convenience, we define
\begin{equation}
\bar{A}=\bigl(A_1\pm\frac{1}{2}A_2\bigr)
\end{equation}
and
\begin{equation}
\bar{B}=\bigl(B_1\pm\frac{1}{2}B_2\bigr)
\end{equation}
where the upper (lower) sign is for the $\alpha$ ($\beta$) phase.

In order to relate the parameters in our model to a physical system, let
us write
\begin{eqnarray}
B_1&=&B_{11}+B_{12}{\tilde T},  \\
B_2&=&B_{21}+B_{22}{\tilde T}.
\end{eqnarray}
Here, $c$ and ${\tilde T}$ are the scaled composition and temperature,
respectively, defined in the text. Let $\Delta C_\nu$ and $m_\nu$ be the
concentration jump at the solid-liquid interface and the liquidus slope of
phase $\nu$, respectively, at $T_p$. Let $r$ be the ratio $\Delta
C_{\beta}/\Delta C_{\alpha}$, then the parameters $A_1$, $A_2$, $B_{11}$,
$B_{12}$, $B_{21}$, and $B_{22}$ are related to these quantities in the
phase diagram by
\begin{eqnarray}
A_1    &=&  \frac{1}{4}(1+r),                         \\
A_2    &=&  \frac{1}{2}(1-r),                         \\
B_{11} &=&  \frac{1}{4}(1+r)\left[r-\frac{1}{4}(1+r)\right],     \\
B_{21} &=&  \frac{1}{2}(1-r)\left[r-\frac{1}{4}(1+r)\right],     \\
B_{12} &=& -\frac{1}{4}\left(1+r\frac{m_\alpha}{m_\beta}\right), \\
B_{22} &=& -\frac{1}{2}\left(1-r\frac{m_\alpha}{m_\beta}\right).    \label{B22}
\end{eqnarray}
In order to have vertical solid-solid coexistence lines, we have to choose
the parameters such as to make $B_{22}$ vanish. The parameters for our
model peritectic system are listed in Table~\ref{simpara}. 
\begin{table}
\caption{Parameters for our model peritectic system.}   \label{simpara}
\begin{center}
\begin{tabular}{ccccc}
\hline
$C_p$         & 38.2~wt~\%         & & $A_1$    & $3.30745\times 10^{-1}$ \\
$C_{p\alpha}$ & 22.1~wt~\%         & & $A_2$    & $3.38509\times 10^{-1}$ \\
$C_{p\beta}$  & 33.0~wt~\%         & & $B_{11}$ & $-2.56790\times 10^{-3}$\\
$m_\alpha$    & $-6.71865$~K/wt~\% & & $B_{21}$ & $-2.62818\times 10^{-3}$\\
$m_\beta$     & $-2.17$~K/wt~\%    & & $B_{12}$ & $-0.5$                  \\
              &                    & & $B_{22}$ & 0.0                     \\
\hline
\end{tabular}
\end{center}
\end{table}

\end{document}